\documentclass[pre,10pt,twocolumn]{revtex4-1}
\usepackage{amssymb}
\usepackage{latexsym}
\usepackage{amsfonts}
\usepackage{graphics}
\usepackage{graphicx}
\usepackage{epsfig}
\usepackage{amsmath}
\usepackage{float}
\usepackage{verbatim}
\usepackage[subfigure]{graphfig}
\usepackage{color}

\begin{document}

\title{How degree distribution broadness influences network robustness:
  comparing localized and random attacks}

\author{Xin Yuan,$^1$ Shuai Shao,$^1$ H. Eugene Stanley,$^1$ and Shlomo
  Havlin$^{1,2}$} 

\affiliation{$^1$Center for Polymer Studies and Department of Physics,
Boston University, Boston, MA 02215 USA\\
$^2$Minerva Center and Department of Physics, Bar-Ilan University,
Ramat-Gan 52900, Israel }

\begin{abstract}

The stability of networks is greatly influenced by their degree
distributions and in particular by their broadness. Networks with
broader degree distributions are usually more robust to random failures
but less robust to localized attacks. To better understand the effect of
the broadness of the degree distribution we study two models in which
the broadness is controlled and compare their robustness against
localized attacks (LA) and random attacks (RA). We study analytically
and by numerical simulations the cases where the degrees in the networks
follow a bi-Poisson distribution $P(k)=\alpha
e^{-\lambda_1}\frac{\lambda_1^k}{k!}+(1-\alpha)
e^{-\lambda_2}\frac{\lambda_2^k}{k!},\alpha\in[0,1]$, and a Gaussian
distribution $P(k)=A \cdot exp{(-\frac{(k-\mu)^2}{2\sigma^2})}$ with a
normalization constant $A$ where $k\geq 0$. In the bi-Poisson
distribution the broadness is controlled by the values of $\alpha$,
$\lambda_1$, and $\lambda_2$, while in the Gaussian distribution it is
controlled by the standard deviation, $\sigma$. We find that only when
$\alpha=0$ or $\alpha=1$, i.e., degrees obeying a pure Poisson
distribution, are LA and RA the same. In all other cases networks are
more vulnerable under LA than under RA. For a Gaussian distribution with
an average degree $\mu$ fixed, we find that when $\sigma^2$ is smaller
than $\mu$ the network is more vulnerable against random
attack. However, when $\sigma^2$ is larger than $\mu$ the network
becomes more vulnerable against localized attack. Similar qualitative
results are also shown for interdependent networks.

\end{abstract}

\pacs{}

\maketitle

\section{Introduction}

Complex networks are widely used as models to understand such features
of complex systems as structure, stability, and function
\cite{watts1998collective,albert2000error,cohen2000resilience,callaway2000network,albert2002statistical,newman2003structure,song2005self,caldarelli2007edited,rosato2008modelling,arenas2008synchronization,
  cohen2010complex,newman2010networks,li2010towards,schneider2011mitigation,bashan2012network,dorogovtsev2013evolution,
  ludescher2013improved,yan2013efficient,boccaletti2014structure,li2015percolation}.
The robustness of networks suffering site or link attacks is a topic of
great interest because it is an important issue affecting many
real-world networks. Such approaches as site percolation on a network
where nodes suffer either random attack (RA)
\citep{albert2000error,cohen2000resilience,callaway2000network} or
targeted attack (TA) based on node connectivity
\cite{albert2000error,cohen2000resilience} have been developed to study
these phenomena. Localized attack (LA) in which nodes surrounding a seed
node are removed layer by layer has also been recently introduced
\cite{shao2015percolation,berezin2015spatially}. In addition,
interdependent networks are more vulnerable to RA and TA than isolated
single networks \cite{buldyrev2010catastrophic, huang2011robustness,
  peixoto2012evolution,baxter2012avalanche,dong2012percolation,bashan2013extreme,radicchi2013abrupt}.
LA on spatially embedded interdependent networks has been addressed, and
a significant metastable regime where LA above a critical size
propagates throughout the whole system has also been found
\cite{berezin2015spatially}.

Although prior research has developed tools for probing network
robustness against all these attack scenarios and has found that degree
distribution broadness strongly influences network stability
\cite{albert2002statistical}, there has been no systematic study of how
degree distribution broadness affects robustness. Here we compare LA and
RA on two networks models in which the broadness is controlled. One
model is bi-Poisson with two groups having different average
degrees. The difference between the two average degrees characterizes
the broadness of the degree distribution of the network. Although
research on this topic usually focuses on a network with a pure Poisson
degree distribution, many real-world networks have two or more degree
distributions \cite{valente2004two,tanizawa2006optimization}. For
example, a network of two groups of people, a high-degree group with
many friends and a low-degree group with few friends, might reflect a
bi-Poisson distribution.  Note that bi-Piossonian networks are optimally
robust against TA \cite{valente2004two}. The second model in which the
broadness can be controlled is a Gaussian degree distribution. Here the
standard deviation $\sigma$ characterizes the broadness of the degree
distribution. This distribution is realistic, e.g., the distribution of
WWW links resembles a Gaussian distribution \cite{pennock2002winners}.

We here analyze the robustness against attack of networks in which we
can tune the broadness of the degree distributions, e.g., those with
bi-Poisson and Gaussian degree distributions.  We limit our approach to
LA and RA and use the frameworks developed in
Refs.~\citep{callaway2000network} and \citep{shao2015percolation},
extending them to study (i) single networks with a bi-Poisson
distribution, (ii) single networks with a Gaussian distribution, (iii)
fully interdependent networks with the same bi-Poisson distribution in
each network, and (iv) fully interdependent networks with the same
Gaussian distribution in each network. By changing $\alpha$ of the
bi-Poisson distribution
\begin{equation}
 P(k)=\alpha e^{-\lambda_1}\frac{\lambda_1^k}{k!}+(1-\alpha)
 e^{-\lambda_2}\frac{\lambda_2^k}{k!},\alpha\in[0,1], 
\end{equation}
with fixed $\lambda_1$ and $\lambda_2$, and $\sigma^{2}$ of the Gaussian
distribution,
\begin{equation}
 P(k)=A \cdot exp{(-\frac{(k-\mu)^2}{2\sigma^2})}, k\geq0,
\end{equation} 
with $\mu$ fixed, we investigate how the distribution broadness
influences the percolation properties. These include the size of
 the giant component $P_{\infty}$ as a
function $p$, the fraction of unremoved nodes and the critical 
threshold $p_c$ at which the giant component $P_{\infty}$ first
collapses. In all cases we find that
our extensive simulations and analytical calculations are in agreement,
and observe the qualitative characteristics of robustness in both single
and interdependent networks under both LA and RA.

\section{RA and LA on a Single Network}

\subsection{Theory}

Following Ref.~\cite{newman2002spread}, we introduce the generating
function of the degree distribution $P(k)$ of a certain network as
\begin{equation}
G_0(x)=\sum_kP(k)x^{k}.
\end{equation}
Similarly, for the generating function of the underlying branching
processes, we have
\begin{equation}
G_1(x)=\sum_k\frac{P(k)k}{\langle k\rangle}x^{k-1}=\frac{G_0^{'}(x)}{G_0^{'}(1)}.
\end{equation}
The size distribution of the clusters that can be reached from a
randomly chosen link is generated in a self-consistent equation
\begin{equation}
H_1(x)=xG_1(H_1(x)).
\end{equation}
Then the size distribution of the clusters that can be traversed by
randomly following a starting vertex is generated by 
\begin{equation}
H_0(x)=xG_0(H_1(x)).
\end{equation}
Next we distinguish between random attack and localized attack.

(\textrm{I}) \textit{Random Attack:} An initial attack with the random
removal of a fraction $1-p$ of nodes from the network changes the
cluster size distribution of the remaining network and the generating
functions of the surviving clusters' size distribution become
\cite{callaway2000network}
\begin{equation}
H_1(x)=1-p+pxG_1(H_1(x)),
\end{equation}
and analogously,
\begin{equation}
H_0(x)=1-p+pxG_0(H_1(x)).
\end{equation}
Here $p_c$, the critical value at which the giant component collapses,
is determined by
\begin{equation}
p_c=\frac{1}{G^{'}_1(1)},
\end{equation}
and
\begin{equation}
p_c=\frac{1}{G^{'}_1(1)}=\frac{G^{'}_0(1)}{G^{''}_0(1)},
\end{equation}
which is equivalent to the expression $p_c={\left\langle k
  \right\rangle}/{\left\langle k(k-1) \right\rangle}$ given in
Ref.~\cite{cohen2000resilience}.

Thus for a bi-Poisson distribution, because $G_0(x)=\alpha
e^{\lambda_1(x-1)}+(1-\alpha) e^{\lambda_2(x-1)}$, $p_c$ is
\begin{equation}
p_c=\frac{\alpha \lambda_1+(1-\alpha) \lambda_2}{\alpha \lambda_1^2+
  (1-\alpha) \lambda_2 ^2}.\label{p_cBi} 
\end{equation}
For a Gaussian distribution we have
\begin{equation}
p_c=\frac{\sum_{1}^{\infty}k
  e^{(-{(k-\mu)^2}/{2\sigma^2})}}{\sum_{2}^{\infty}k(k-1)
  e^{(-{(k-\mu)^2}/{2\sigma^2})}}.\label{p_cGau} 
\end{equation}

The size of the resultant giant component is \cite{callaway2000network}
\begin{equation}
P_\infty(p)=1-H_0(1)=p[1-G_0(H_1(1))],\label{GRA}
\end{equation}
which can be numerically determined by solving $H_1(1)$ from its
self-consistent equation
\begin{equation}
H_1(1)=1-p+pG_1(H_1(1)).
\end{equation}

(\textrm{II}) \textit{Localized Attack}: We next consider the local
removal of a fraction $1-p$ of nodes, starting with a randomly chosen
seed node. Here we remove the seed node and its nearest neighbors,
next-nearest neighbors, next-next-nearest neighbors, and continue until
a fraction $1-p$ of nodes have been removed from the network. This
pattern of attack reflects such real-world cases as earthquakes or the
use of weapons of mass destruction.  As in
Ref.~\citep{shao2015percolation}, the localized attack occurs in two
stages, (i) nodes belonging to the attacked area (the seed node and the
layers surrounding it) are removed but the links connecting them to the
remaining nodes of the network are left in place, but then (ii) these
links are also removed. Following the method introduced in
Refs.~\cite{shao2015percolation,shao2009structure}, we find the
generating function of the degree distribution of the remaining network to be
\begin{equation}
G_{p0}(x)=\frac{1}{G_0(f)}G_0[f+\frac{G^{'}_0(f)}{G^{'}_0(1)}(x-1)],\label{Gp0}
\end{equation}
where $f\equiv G^{-1}_0(p)$.  The generating function of the
underlying branching process is thus
\begin{equation}
G_{p1}(x)=\frac{G^{'}_{p0}(x)}{G^{'}_{p0}(1)}.
\end{equation}
The generating function of the cluster size distribution following a
random starting node in the remaining network is
\begin{equation}
H_{p0}(x)=xG_{p0}(H_{p1}(x)),
\end{equation}
where $H_{p1}(x)$, the generating function of the cluster size
distribution given by randomly traversing a link, satisfies the
self-consistent condition
\begin{equation}
H_{p1}(x)=xG_{p1}(H_{p1}(x)). \label{Hp1}
\end{equation}
The network begins to generate a giant component when $G^{'}_{p1}(1)=1$
\citep{shao2015percolation}, which yields $p_c$ as the solution to
\begin{equation}
G^{''}_0(G^{-1}_0(p_c))=G^{'}_0(1).\label{LAC}
\end{equation}
The size of the giant component $P_\infty(p)$ as a fraction of the
remaining network thus satisfies \citep{shao2015percolation}
\begin{equation}
P_\infty(p)=p\left[1-G_{p0}(H_{p1}(1))\right],\label{GLA}
\end{equation}
which can be numerically determined by first solving $H_{p1}(1)$ from
Eq.~(\ref{Hp1}), i.e., $H_{p1}(1)=G_{p1}(H_{p1}(1))$.

\begin{figure}
\includegraphics[width=0.48\textwidth, angle = 0]{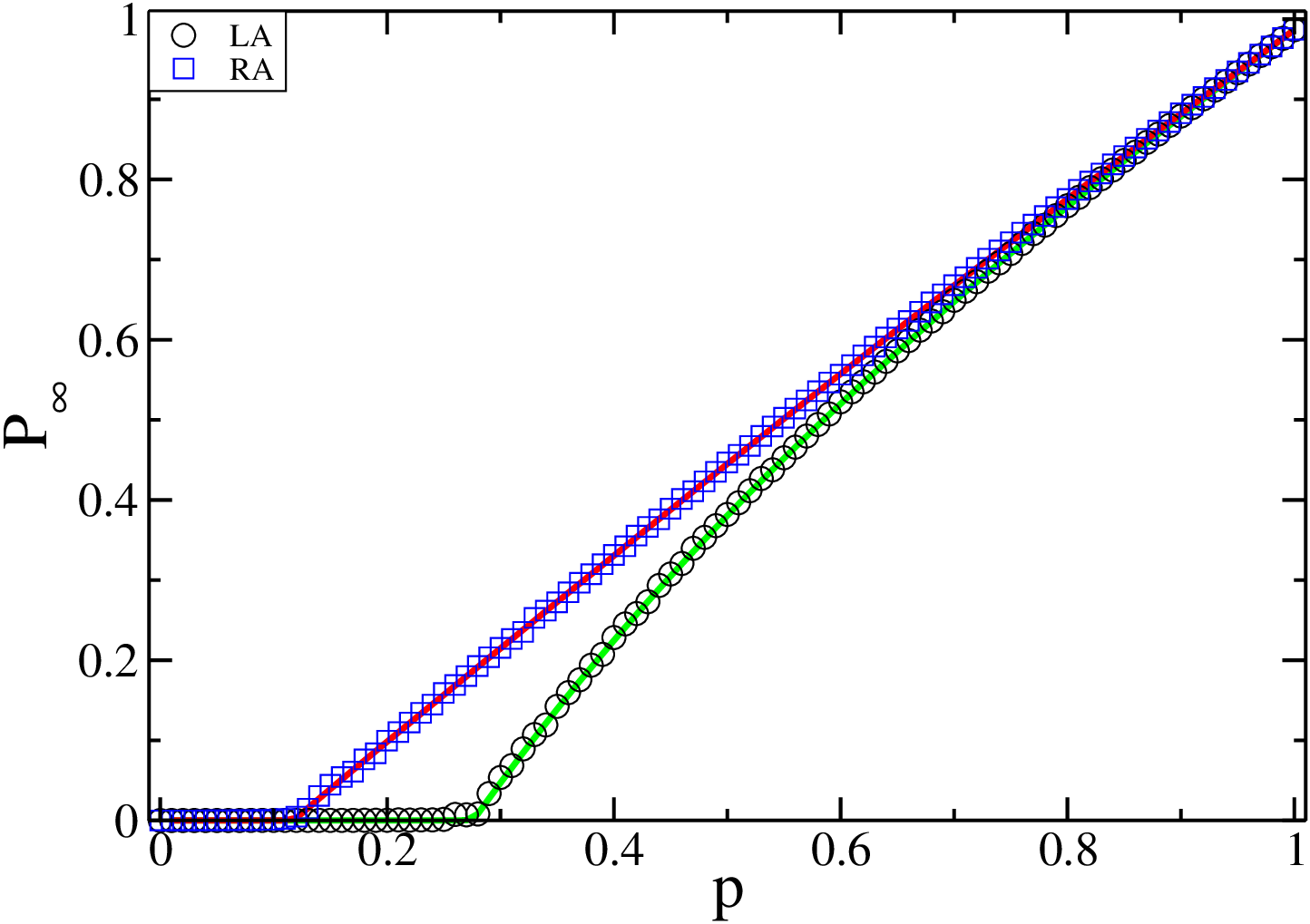}
\caption{\label{Fig1}(Color online) Sizes of giant component,
  $P_{\infty}(p)$, as a function of $p$ for $\lambda_1=4$,
  $\lambda_2=12$ and $\alpha=0.7$. Here solid lines are theoretical
  predictions, from Eq.~(\ref{GRA}) for RA (red line) and
  Eq.~(\ref{GLA}) for LA (green line), and symbols are simulation
  results with network size $N=10^4$, where averages are taken over 10
  realizations, under LA ($\bigcirc$) and RA ($\square$). }
\end{figure}
 
In order to determine $p_c$ explicitly, we first get $f_c$ from
$f_c\equiv G^{-1}_0(p_c)$, i.e., $f_c$ from $G_0(f_c)=p_c$. Then from
Eq.~(\ref{LAC}) $f_c$ must also satisfy $G^{''}_0(f_c)=G^{'}_0(1)$. In
the general case, $p_c$ and $P_{\infty}$ must be obtained by solving
numerically Eqs.~(\ref{LAC}) and (\ref{GLA}).  In certain limiting
cases, however, one can derive explicit analytical expressions for $p_c$
that yield more physical insight. An example of a specific case is given
in the next subsection.

\subsubsection{Analytic solution of $p_c$ for bi-Poisson distribution
  with $\lambda_2 =2 \lambda_1$} 

For a bi-Poisson distribution, using its generating function and
$G_0(f_c)=p_c$, $f_c$ and $p_c$ satisfy the relation
\begin{equation}
G_0(f_c)=\alpha [e^{(f_c-1)}]^{\lambda_1}+(1-\alpha)
[e^{(f_c-1)}]^{\lambda_2}=p_c.\label{f_c1} 
\end{equation} 
Assuming $\lambda_2 =2 \lambda_1$, we denote $e^{\lambda_1(f_c-1)}=y$
such that Eq.~(\ref{f_c1}) reduces to $\alpha y+(1-\alpha) y^{2}=p_c$,
which, for $\alpha\neq 1$, is a quadratic equation of $y$ and its
positive solution is
\begin{equation}
y=\frac{\sqrt{\alpha^2+ 4p_c(1-\alpha)}-\alpha}{2(1-\alpha)}.\label{y1}
\end{equation}
Plugging $f_c$ into Eq.~(\ref{LAC}) we get another quadratic equation of
$y$,
\begin{equation}
\alpha \lambda_1^2y+(1-\alpha)\lambda_2^2y^2=\alpha \lambda_1+(1-\alpha)\lambda_2,
\end{equation}
for which the physical solution of $y$ is 
\begin{equation}
y=\frac{\sqrt{\alpha^2 \lambda_1^4+4(1-\alpha)\lambda_2^2[\alpha
      (\lambda_1-\lambda_2)+\lambda_2]}-\alpha  \lambda_1^2}{2(1-\alpha)\lambda_2^2}. \label{y2}
\end{equation}
Because $f_c=ln(y)/\lambda_1+1$, to obtain $p_c$ we need to equate
Eqs.~(\ref{y1}) and (\ref{y2}). Thus we obtain
\begin{equation}
p_c=\frac{1}{64(1-\alpha)}
[\beta+6\alpha\sqrt{\alpha^2+\beta}-6\alpha^2],\label{p_cLABi} 
\end{equation}
where $\beta=\frac{16(1-\alpha)(2-\alpha)}{\lambda_1}$. We use the
relation of $\lambda_2 =2 \lambda_1$ for simplification. Plugging
$\alpha=0$ into Eq.~(\ref{p_cLABi}), we get $p_c={1}/{\lambda_2}$ as
found in Ref.~\citep{shao2015percolation}. For $\alpha\rightarrow1$,
employing the L'H\^{o}pital rule we also get $\lim_{\alpha \to
  1}p_c={1}/{\lambda_1}$, as found in the pure Poisson distribution
described above.

\begin{figure}
 \includegraphics[width=0.48\textwidth, angle = 0]{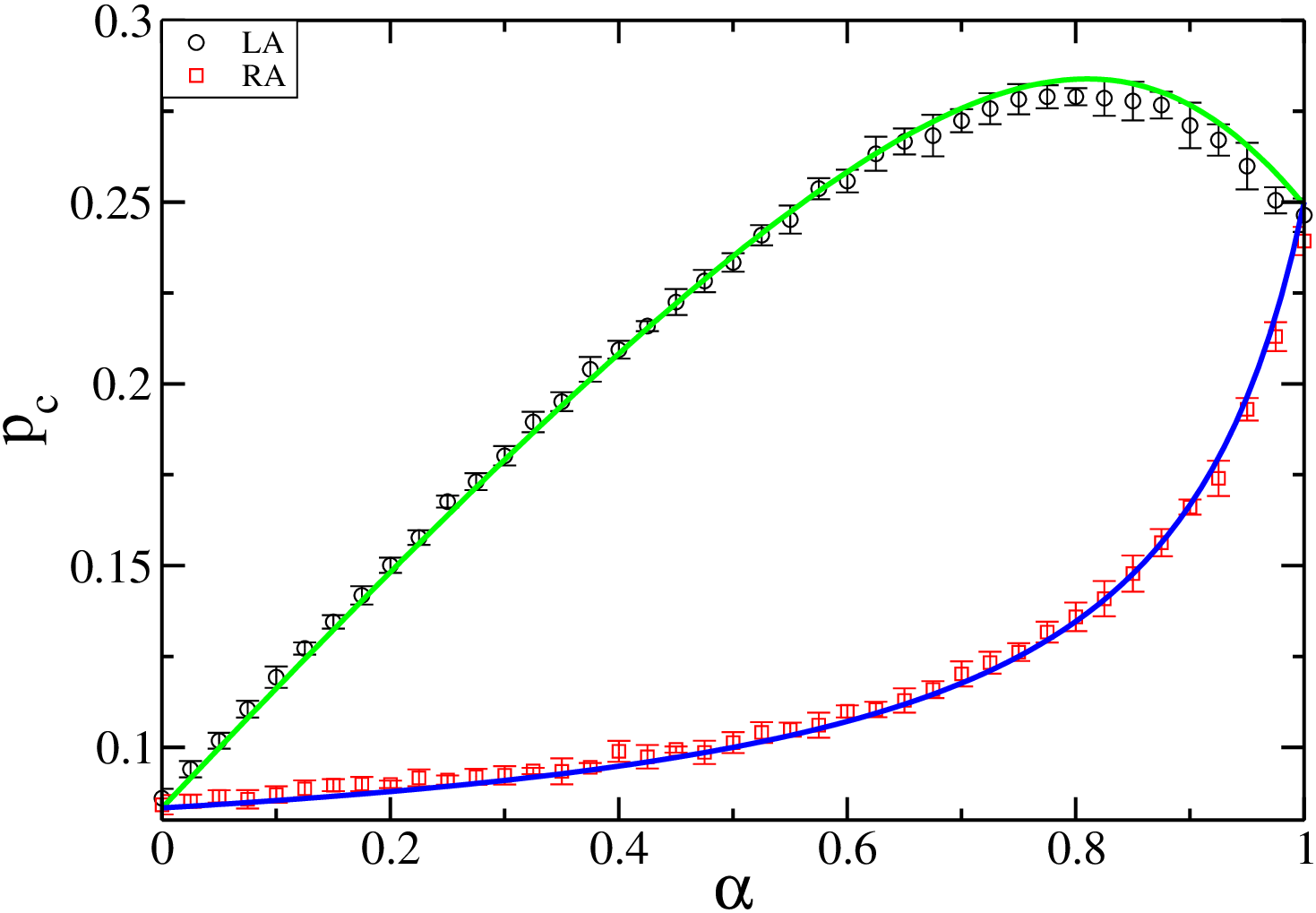}
\caption{\label{Fig2}(Color online) Percolation thresholds $p_c$ of a
  single bi-Poisson network as a function of $\alpha$ under LA and RA
  with $\lambda_1=4$, $\lambda_2=12$.  Here solid lines are theoretical
  predictions, from Eq.~(\ref{p_cBi}) for RA (blue line) and
  Eq.~(\ref{LAC}) for LA (green line) and symbols ($\square$ for RA and
  $\bigcirc$ for LA) with error bars are simulation results with network
  size of $N=10^4$ nodes, where averages and standard deviations are
  taken over 20 realizations.}
\end{figure}

It is impossible to derive $p_c$ explicitly for a Gaussian distribution.
Even for a bi-Poisson distribution, other than special cases such as the
one discussed above, deriving $p_c$ is also impossible because it
requires solving first $f_c=G^{-1}_0(p_c)$, i.e., $f_c$ from
Eq.~(\ref{f_c1}), which could be viewed as $\alpha
y^{\lambda_1}+(1-\alpha) y^{\lambda_2}=p_c$, a polynomial equation of
$y=e^{(f_c-1)}$. Because we also consider the cases of $\lambda_2 >
\lambda_1 \geqslant 4$ using the Abel-Ruffini theorem, there is no
general algebraic solution to the above equation except in some special
cases.  Hence we use the Newton method to solve $p_c$ and $P_{\infty}$
numerically.

\subsection{Results}

To test the analytical predictions above we conduct numerical
calculations of analytic expressions, and compare the results with the
simulation results on single networks with degrees following both
bi-Poisson distributions and Gaussian distributions under both LA and
RA. All the simulation results are obtained for networks of $N=10^4$
nodes.

\subsubsection{Single bi-Poisson networks}

Figure~\ref{Fig1} shows the giant component $P_{\infty}(p)$ as a
function of the occupation probability $p$ under LA and RA. Note that
$p_c$ is larger for LA than for RA. The simulation results agree with
the theoretical results obtained from Eqs.~(\ref{GRA}) and (\ref{GLA}),
and there is second-order percolation transition behavior in both attack
scenarios. Note that when $\alpha= 0$ or $1$, i.e., when node degrees
follow a pure Poisson distribution as reported in Ref.~\cite{shao2015percolation},
the networks have the same critical value of
$p_c$ under LA and RA and the same dependence of $P_{\infty}(p)$ on
$p$. However when $\alpha=0.7$, $p_c(LA)>p_c(RA)$, indicating that the
network is more fragile under LA than under RA, and that the giant
components exhibit different behavior.

\begin{figure}
\includegraphics[width=0.48\textwidth, angle = 0]{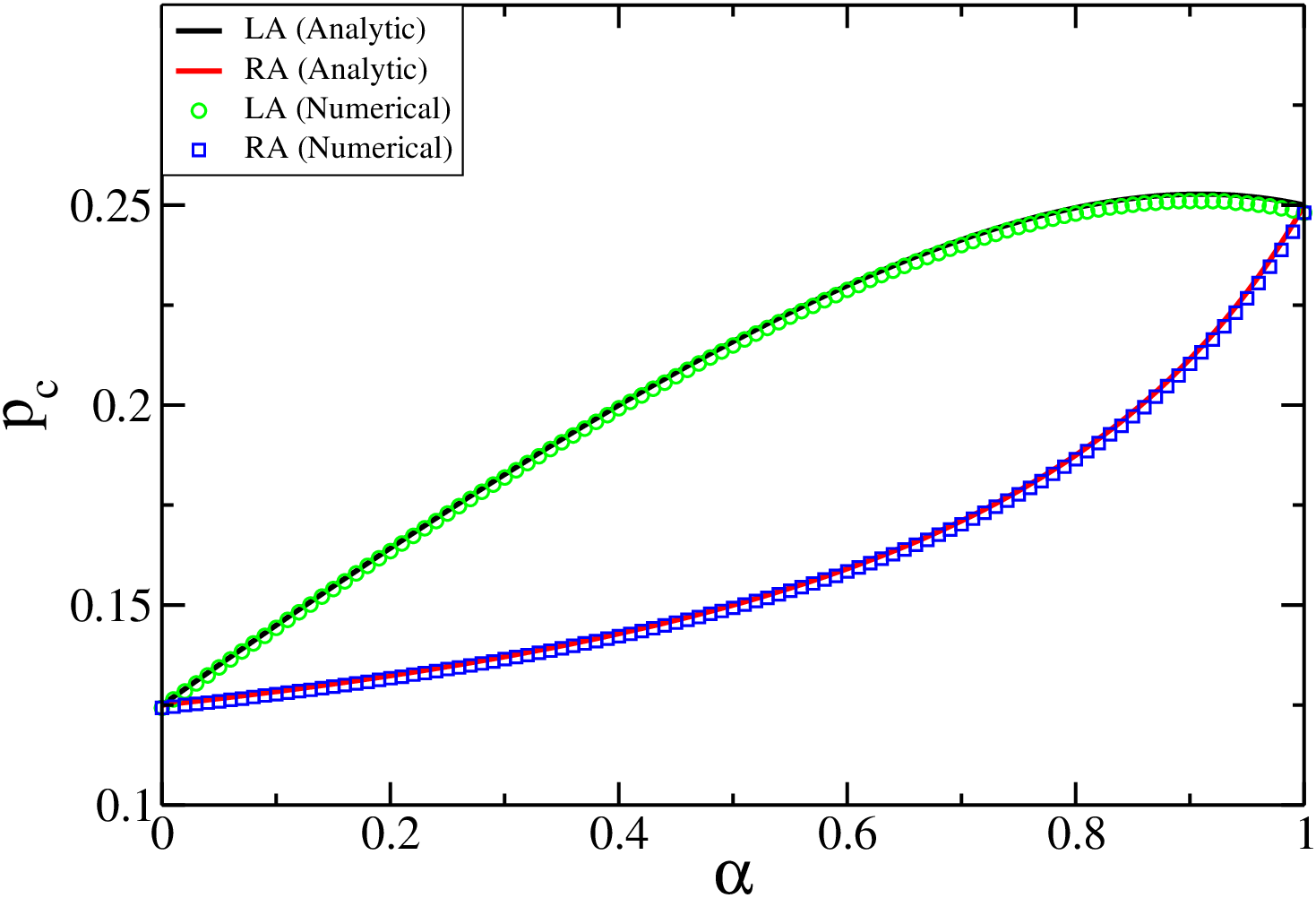}
\caption{(Color online)\label{Fig3} Comparison between numerical results
  (symbols) and the analytic results (solid lines) for bi-Poisson
  distribution with $\lambda_1=4$ and $\lambda_2=8$. Note that they
  agree with each other well. Here, all the analytic results are
  obtained from Eq.~(\ref{p_cBi}) for RA (red line) and
  Eq.~(\ref{p_cLABi}) for LA (black line) and the numerical results are
  attained by employing Newton's Method on Eqs.~(\ref{p_cBi}) and
  (\ref{LAC}) respectively. }
\end{figure}

Figure ~\ref{Fig2} shows how the broadness of the distribution, tuned by
changing $\alpha$ with fixed $\lambda_1$ and $\lambda_2$, influences the
robustness of the network under LA and RA. The solid lines are the
numerical results obtained from the Newton method and the symbols with
error bars are the simulation results. Note that only when $\alpha=0$
and $\alpha=1$ does $p_c(LA)=p_c(RA)$. In all other cases
$p_c(LA)>p_c(RA)$, indicating that the network is always more vulnerable
under LA than under RA if the degree distribution is bi-Poissonian. Note
also that $p_c(LA)$ peaks at $\alpha=0.79$.

\begin{figure}
\includegraphics[width=0.48\textwidth, angle = 0] {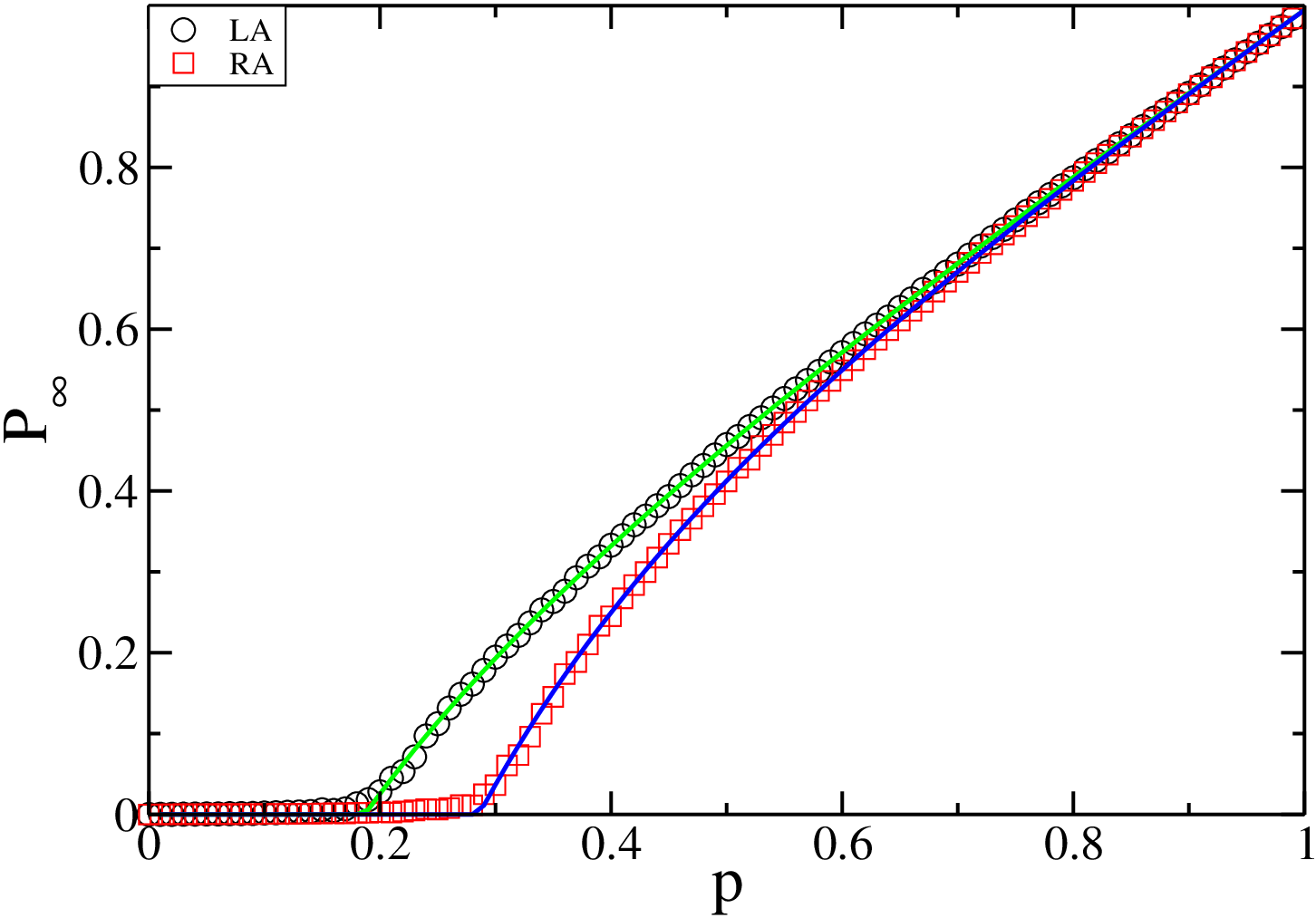}
\caption{\label{Fig4}(Color online) Sizes of giant component as a
  function of $p$ of a single Gaussian network with $\mu=4$ and
  $\sigma^2=2$.  Here solid lines are theoretical results, from
  Eq.~(\ref{GRA}) for RA (blue line) and Eq.~(\ref{GLA}) for LA (green
  line) and symbols are simulation results obtained from network size of
  $N=10^4$ where averages are taken over 10 realizations under LA
  ($\bigcirc$) and RA ($\square$).}
\end{figure}

For the special case of $\lambda_2=2 \lambda_1$, we compare the
analytical values of $p_c$ from Eqs.~(\ref{p_cBi}) and (\ref{p_cLABi})
using $\lambda_1=4$ and $\lambda_2=8$ with results obtained from the
Newton method (see Fig.~\ref{Fig3}). For this combination of average
degrees, $p_c(LA)$ peaks at $\alpha=0.91$. Note that the results agree,
indicating that the Newton method produces satisfactory results and
therefore, in the general case in which $\lambda_2 \neq 2 \lambda_1$ and
in the cases of Gaussian distribution, it can be used to get $p_c(LA)$.

\subsubsection{Single Gaussian Networks}
 
Figure~\ref{Fig4} shows the giant component $P_{\infty}(p)$ as a
function of the occupation probability $p$ under LA and RA respectively
for a single network with a Gaussian degree distribution. Note that the
simulation results and the theoretical results obtained from
Eqs.~(\ref{GRA}) and (\ref{GLA}) agree, and that second-order phase
transition behavior is present in both attack scenarios. Note also that
$\mu=4$ and $\sigma^2=2$, and thus $p_c(LA)<p_c(RA)$, which indicates
that the network is more robust under LA than under RA for this
particular distribution.

\begin{figure}
\includegraphics[width=0.48\textwidth,angle=0]{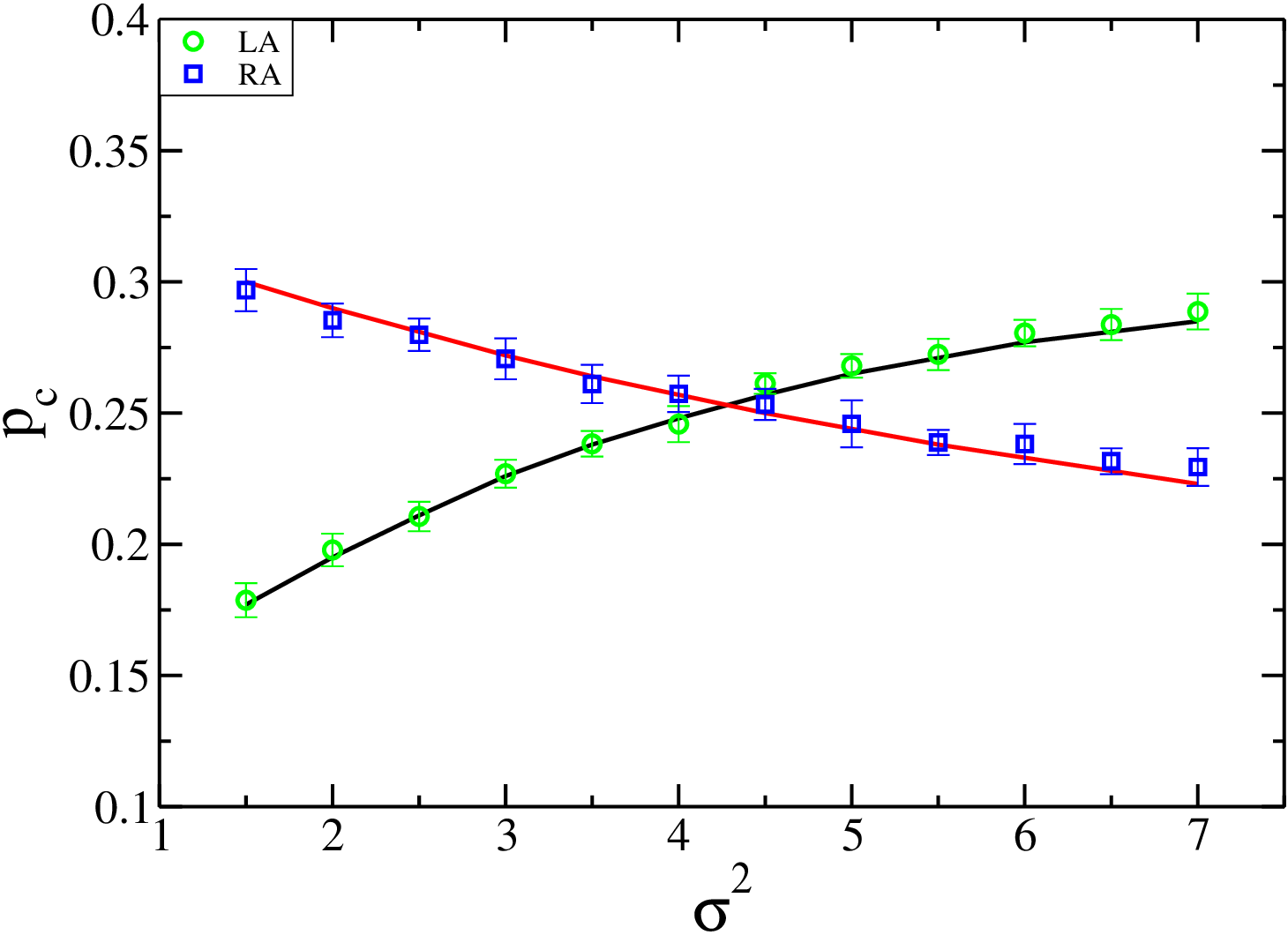}
\caption{\label{Fig5}(Color online) Percolation thresholds $p_c$ as a
  function of $\sigma^2$ of networks with Gaussian degree distribution
  under LA and RA with $\mu=4$.  Here solid lines are theoretical
  predictions, from Eq.~(\ref{p_cGau}) for RA (red line) and
  Eq.~(\ref{LAC}) for LA (black line) and symbols ($\square$ for RA and
  $\bigcirc$ for LA) with error bars are simulation results with network
  size of $N=10^4$ nodes, where averages and standard deviations are
  taken over 20 realizations. It is shown here that as $\sigma^2$
  increases $p_c(LA)$ increases whereas $p_c(RA)$ decreases
  simultaneously and they intersect each other around $\sigma^2 \approx
  \mu=4$. }
\end{figure}

We fix $\mu$ and find that when the Gaussian distribution gets broader,
i.e., when $\sigma$ increases, $p_c(RA)$ decreases, but that $p_c(LA)$
increases with $\sigma$ (see Fig.~\ref{Fig5}). Note that when
$\sigma^2<\mu$, $p_c(LA)<p_c(RA)$, and that the opposite is true when
$\sigma^2>\mu$. Note also that when $\sigma^2 \approx \mu$ there is a
crossing point with $p_c(RA)\approx p_c(LA)$, which is analogous to a
Poisson ER network with the same mean and variance and the robustness of
the network under both LA and RA is the same, as reported in
Ref.~\cite{shao2015percolation}.

\begin{figure}
\includegraphics[width=0.48\textwidth, angle = 0]{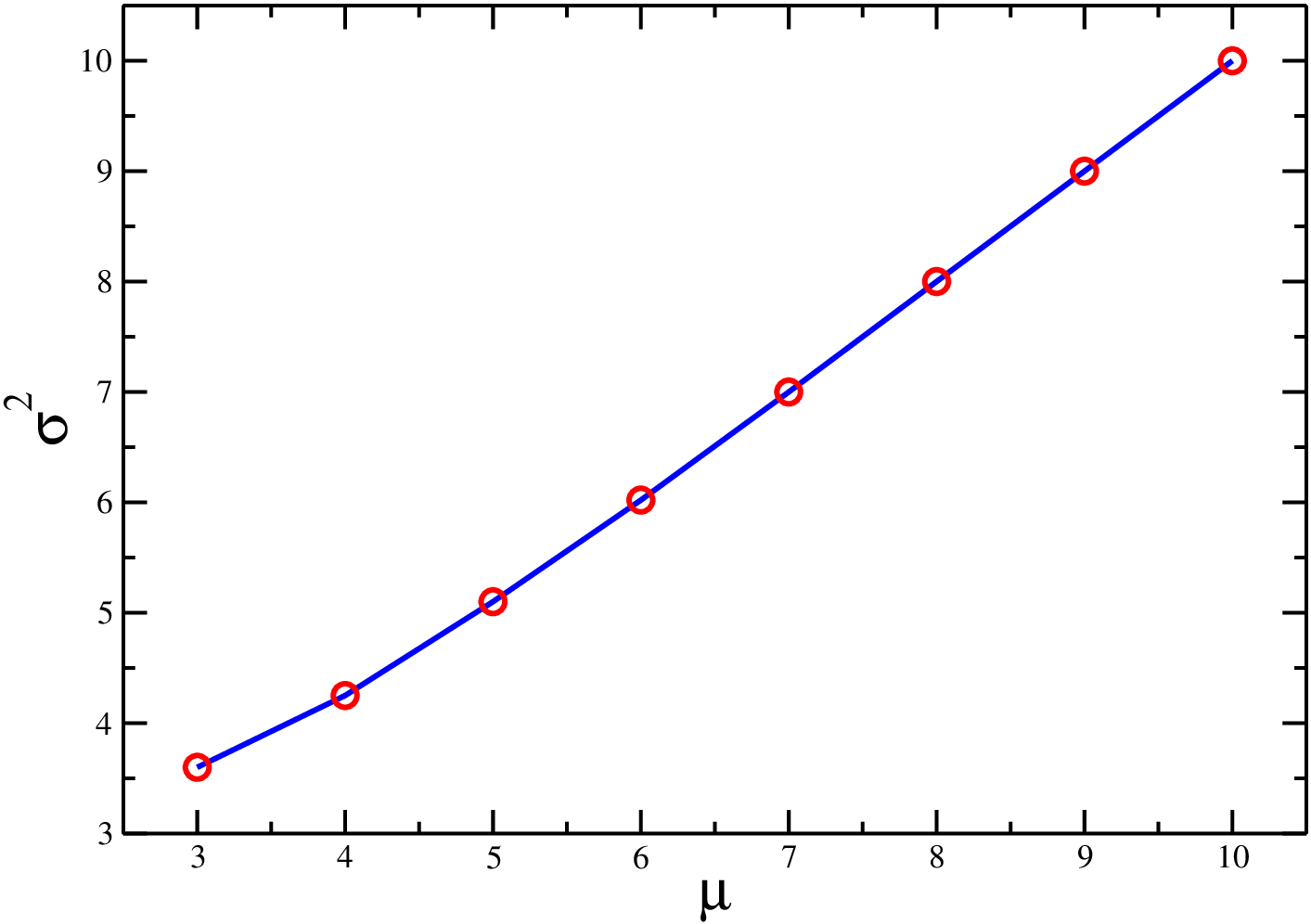}
\caption{(Color online)\label{Fig6} $\sigma^2$ as a function of $\mu$
  at the intersection point where $p_c(LA)=p_c(RA)$ for single networks
  where degrees follow a Gaussian distribution.}
\end{figure}

Figure~\ref{Fig6} shows a plot of $\sigma^2$ as a function of $\mu$
when this intersection point occurs, i.e., when $p_c(LA)=p_c(RA)$. Note
that except for some minor deviations at small $\mu$ values, because
$k\geq 0$ the Gaussian distribution is deformed, the region above the
extrapolation curve corresponds to $p_c(LA)>p_c(RA)$, and the region
below to $p_c(LA)<p_c(RA)$.

\section{RA and LA on Fully Interdependent Networks}

\subsection{Theory}

We apply the formalism of RA on fully interdependent networks introduced
in Ref.~\cite{buldyrev2010catastrophic}. Specifically, we consider two
networks $A$ and $B$ with the same number of nodes $N$. Within each
network the nodes are randomly connected with degree distributions
$P_A(k)$ and $P_B(k)$ respectively. Every node in network $A$ depends on
a random node in network $B$, and vice versa.  We also assume that if a
node $i$ in network $A$ depends on a node $j$ in network $B$ and node
$j$ depends on node $l$ in network $A$, then $l=i$, which rules out the
feedback condition \cite{gao2013percolation}. This full interdependency
means that every node $i$ in network $A$ has a dependent node $j$ in
network $B$, and if node $i$ fails node $j$ will also fail, and vice
versa.

\begin{figure}
 \includegraphics[width=0.48\textwidth, angle = 0]{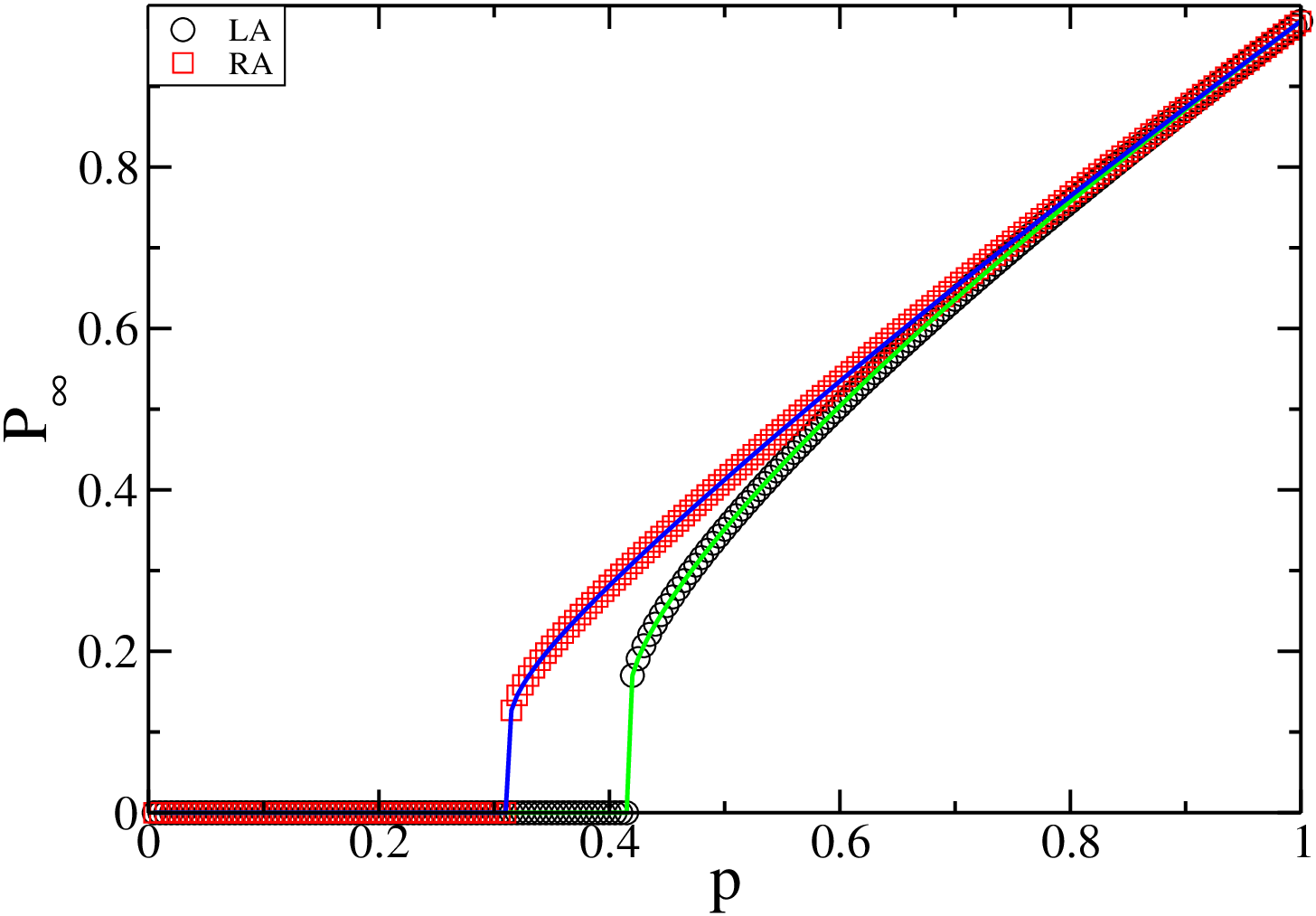}
\caption{\label{Fig7}(Color online) Sizes of the mutually connected
  giant component of the fully interdependent bi-Poisson networks as a
  function of $p$ for $\lambda_1=4$, $\lambda_2=12$ and
  $\alpha=0.5$. Here solid lines are theoretical predictions, from
  Eq.~(\ref{MCGG}) for RA (blue line) and similarly for LA (green line),
  and symbols are simulation results with network size $N=10^4$, where
  averages are taken over 10 realizations, under LA ($\bigcirc$) and RA
  ($\square$). }
\end{figure}

(I) \textit{Random Attack}: We begin by randomly removing a fraction
$1-p$ of nodes and their links in network $A$. All the nodes in network
$B$ that are dependent on the removed nodes in network $A$ are also
removed along with their connectivity links. As nodes and links are
sequentially removed, each network begin to break down into connected
components. Due to interdependency, the removal process iterates back
and forth between the two networks until they fragment completely or
produce a mutually connected giant component with no further
disintegration.  As in Ref.~\cite{buldyrev2010catastrophic} we introduce
the function $g_A(p)=1-G_{A0}[1-p(1-f_A)]$, which is the fraction of
nodes that belong to the giant component of network $A$, where $f_A$ is
a function of $p$ that satisfies the transcendental equation
$f_A=G_{A1}[1-p(1-f_A)]$. Similar equations exist for network $B$. When
the system of interdependent networks stops disintegrating, the fraction
of nodes in the mutual giant component is $P_\infty$, satisfying
\begin{equation}
P_\infty=xg_B(x)=yg_A(y), \label{MCGG}
\end{equation}
where $x$ and $y$ satisfy 
\begin{equation}
x=pg_A(y),    y=pg_B(x).\label{xy}
\end{equation}

Excluding the trivial solution $x=0, y=0$ to the equation set above, we
combine them into a single equation by substitution and obtain,
\begin{equation}
x=g_A[g_B(x)p]p.\label{x}
\end{equation}

\begin{figure}
\includegraphics[width=0.48\textwidth,angle=0]{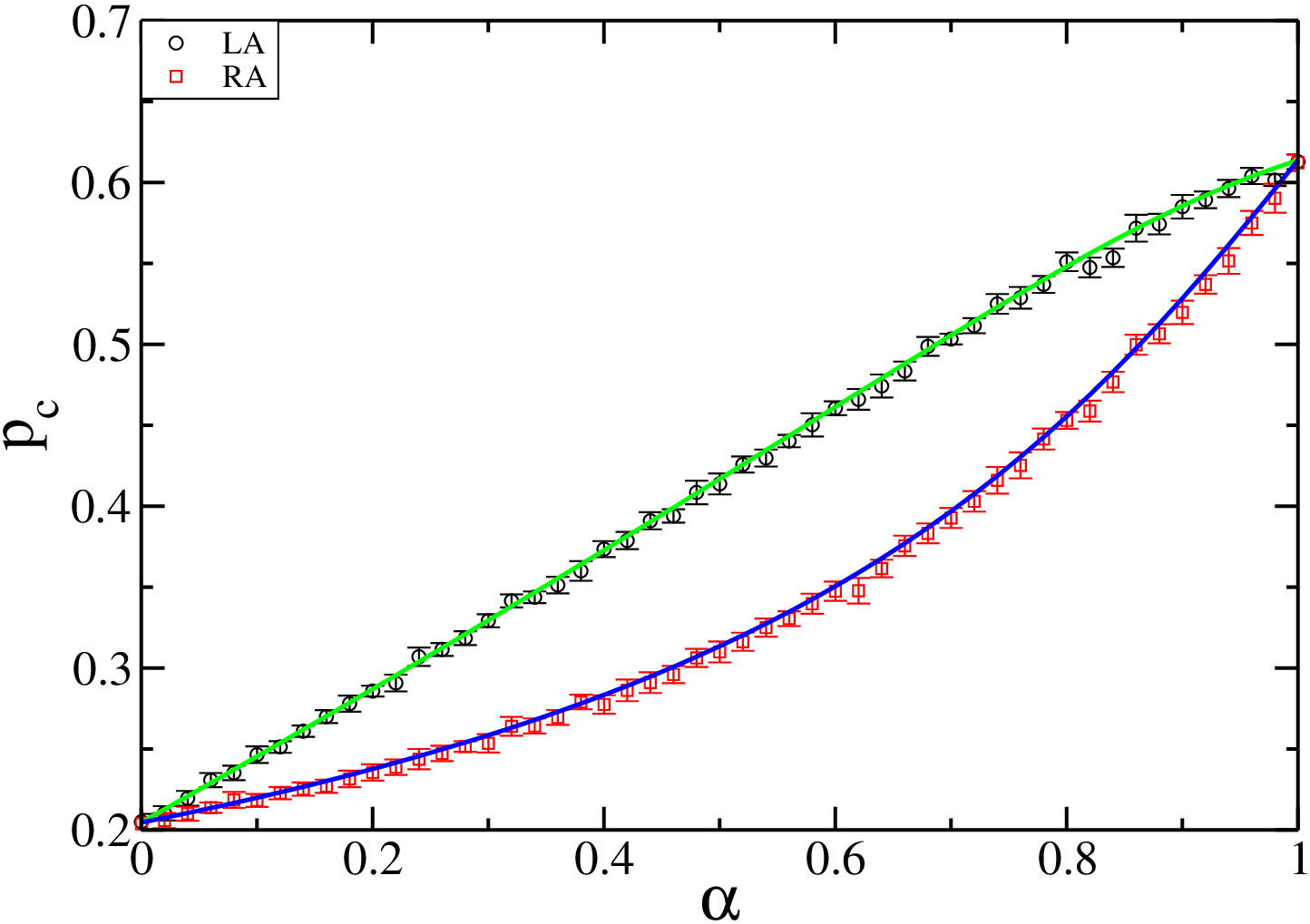}
\caption{\label{Fig8}(Color online) Percolation thresholds $p_c$ of the
  fully interdependent bi-Poisson networks with $\lambda_1=4$,
  $\lambda_2=12$ as a function of $\alpha$ under LA and RA.  Here solid
  lines are theoretical predictions, from Eq.~(\ref{tp_c}) for RA (blue
  line) and similarly for LA (green line) and symbols ($\square$ for RA
  and $\bigcirc$ for LA) with error bars are simulation results with
  network size of $N=10^4$ nodes, where averages and standard deviations
  are taken from 20 realizations. When $\alpha$ is not 1 or 0, $p_c(LA)$
  is always larger than $p_c(RA)$.}
\end{figure}

A nontrivial solution emerges in the critical case ($x=x_c,p=p_c$) by
equating the derivatives of both sides of Eq.~(\ref{x}) with respect to
$x$
\begin{equation}
1=p^2 \frac{dg_A[pg_B(x)]}{dx}\frac{dg_B(x)}{dx}\vline _{x=x_c,p=p_c} \label{tp_c}
\end{equation}
which, together with Eq.~(\ref{xy}), gives the solution for $p_c$ and
the critical size of the mutually connected component,
$P_\infty(p_c)=x_cg_B(x_c)$.

(II) \textit{Localized Attack}: When LA is performed on the one-to-one
fully interdependent networks $A$ and $B$ described above, we can find
an equivalent random network $E$ with generating function $G_{E0}(x)$
such that after a random attack in which $1-p$ nodes in network $E$ are
removed, the generating function of the degree distribution of the remaining
network is the same as $G_{p0}(x)$ (with the substitution of $G_0(x)$ by
$G_{A0}(x)$). Then the LA problem on networks $A$ and $B$ can be 
mapped to a RA problem on networks $E$ and $B$. By using $G_{E0}(1-p+px)
=G_{p0}(x)$ and from Eq.~(\ref{Gp0}) we have
\begin{equation}
G_{E0}(x)=\frac{1}{G_{A0}(f)}G_{A0}
[f+\frac{G^{'}_{A0}(f)}{G^{'}_{A0}(1)G_{A0}{(f)}}(x-1)].\label{equiv} 
\end{equation}

Thus by mapping the LA problem on interdependent networks $A$ and $B$
to a RA problem on a transformed pair of interdependent networks $E$ and
$B$, we can apply the mechanism of RA on interdependent networks to
solve $p_c$ and $P_\infty(p)$ under LA.

\begin{figure}
\includegraphics[width=0.48\textwidth, angle = 0] {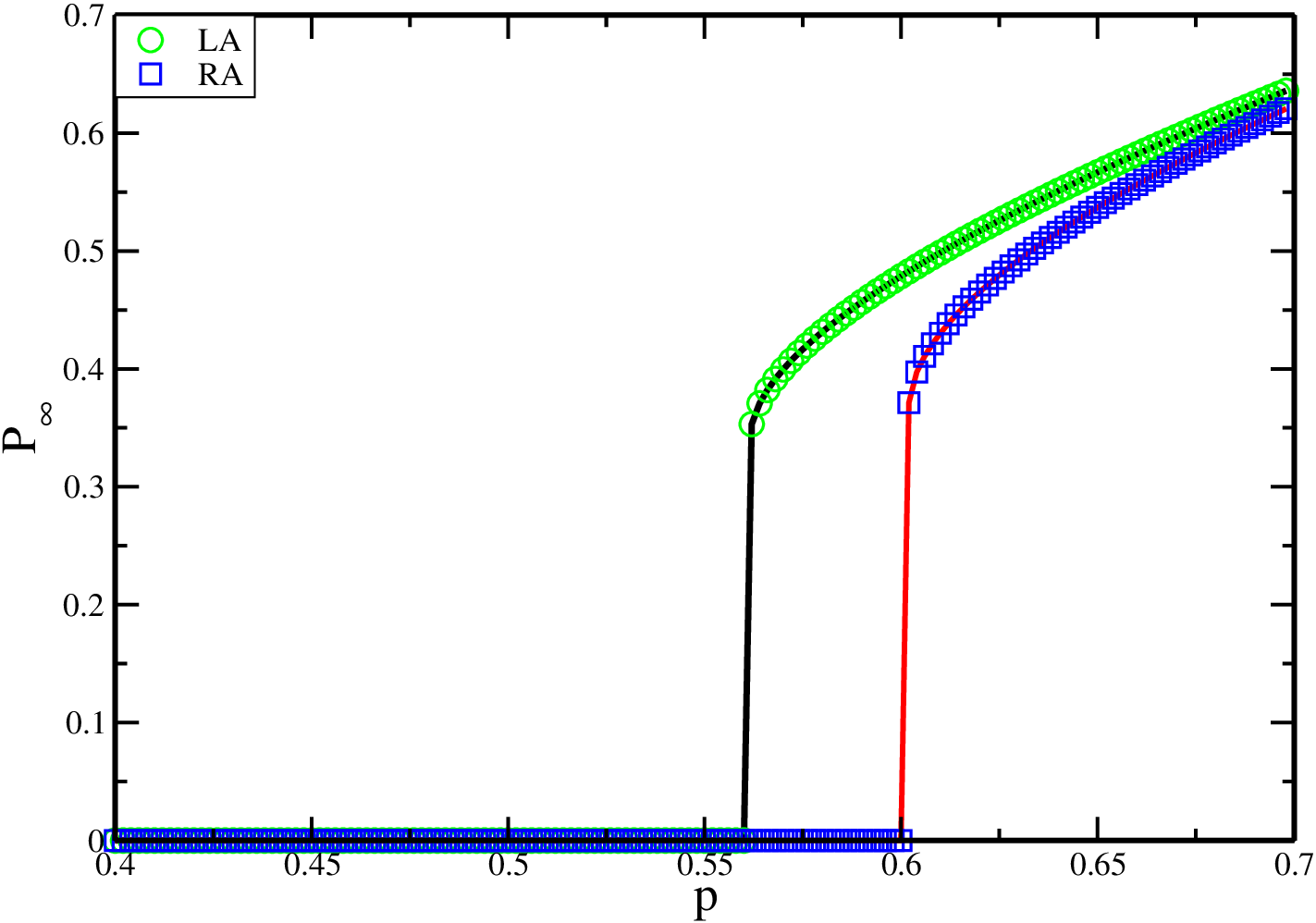}
\caption{\label{Fig9}(Color online) Sizes of mutual giant component of
  the fully interdependent Gaussian networks as a function of $p$ with
  $\mu=4$ and $\sigma^2=2$. Here solid lines are theoretical
  predictions, from Eq.~(\ref{MCGG}) for RA (red line)and similarly for
  LA (black line), and symbols are simulation results with network size
  $N=10^4$, where averages are taken over 10 realizations, under LA
  ($\bigcirc$) and RA ($\square$).}
\end{figure}

Note that for pure Poisson distributions, $f\equiv
G^{-1}_{A0}(p)=\frac{ln(p)}{\lambda}+1$, and that by substituting $f$
into Eq.~(\ref{equiv}) we get $G_{E0}(x)=G_{A0}(x)$. Thus we find that
pure Poisson distributions have exactly the same percolation properties
for fully interdependent networks under LA as those under RA, as found
in Ref.~\citep{buldyrev2010catastrophic}. Because the extreme complexity of
the above equations makes it difficult to obtain explicit expressions for
$p_c$ and $P_\infty(p)$ except when degree distributions are simple, we
resort to numerical calculations in general.

\subsection{Results}

\subsubsection{Fully interdependent networks with bi-Poisson degree distribution }

We start with two fully interdependent networks in which the degrees
both follow the same bi-Poisson distribution and carry out a RA on one
of the networks, initiating a cascading failure process that will
continue until equilibrium is reached. We then do the same procedure
with the same set-up but this time using a LA to initiate the cascading
failure process.  Figure~\ref{Fig7} shows the size of the giant
component $P_{\infty}(p)$ of the system as a function of the occupation
probability $p$ under LA and under RA. Note that in both RA and LA
scenarios the simulation results and the theoretical results obtained
from Eq.~(\ref{MCGG}) agree, indicating that our strategy of finding an
equivalent network under LA works. The first-order phase transition that
occurs in both attack scenarios indicates that the interdependency of
the system makes it much more vulnerable to attack than single
networks. When $\alpha=0.5$ the system is more fragile under LA than
under RA with $p_c(LA)> p_c(RA)$, and the giant components exhibit
different behaviors.

Figure~\ref{Fig8} shows how the broadness of the distribution, tuned by
changing $\alpha$ with fixed $\lambda_1$ and $\lambda_2$, influences the
robustness of the network under both LA and RA. Solid lines are
numerical results using the Newton method on Eq.~(\ref{tp_c}) and
symbols with error bars are simulation results. Note that only when
$\alpha=0$ and $\alpha=1$ is $P(k)$ reduced to a pure Poisson, and we
have $p_c(LA)=p_c(RA)=2.4554/\left\langle k\right\rangle$, as in
Ref.~\cite{buldyrev2010catastrophic}. When $\alpha$ deviates from 0 or
1, i.e., when $P(k)$ deviates from a pure Poisson distribution and takes
the form of a bi-Poisson distribution, $p_c(LA)>p_c(RA)$, indicating
that the system is more vulnerable under LA than under RA.

\begin{figure}
\includegraphics[width=0.48\textwidth,angle=0]
{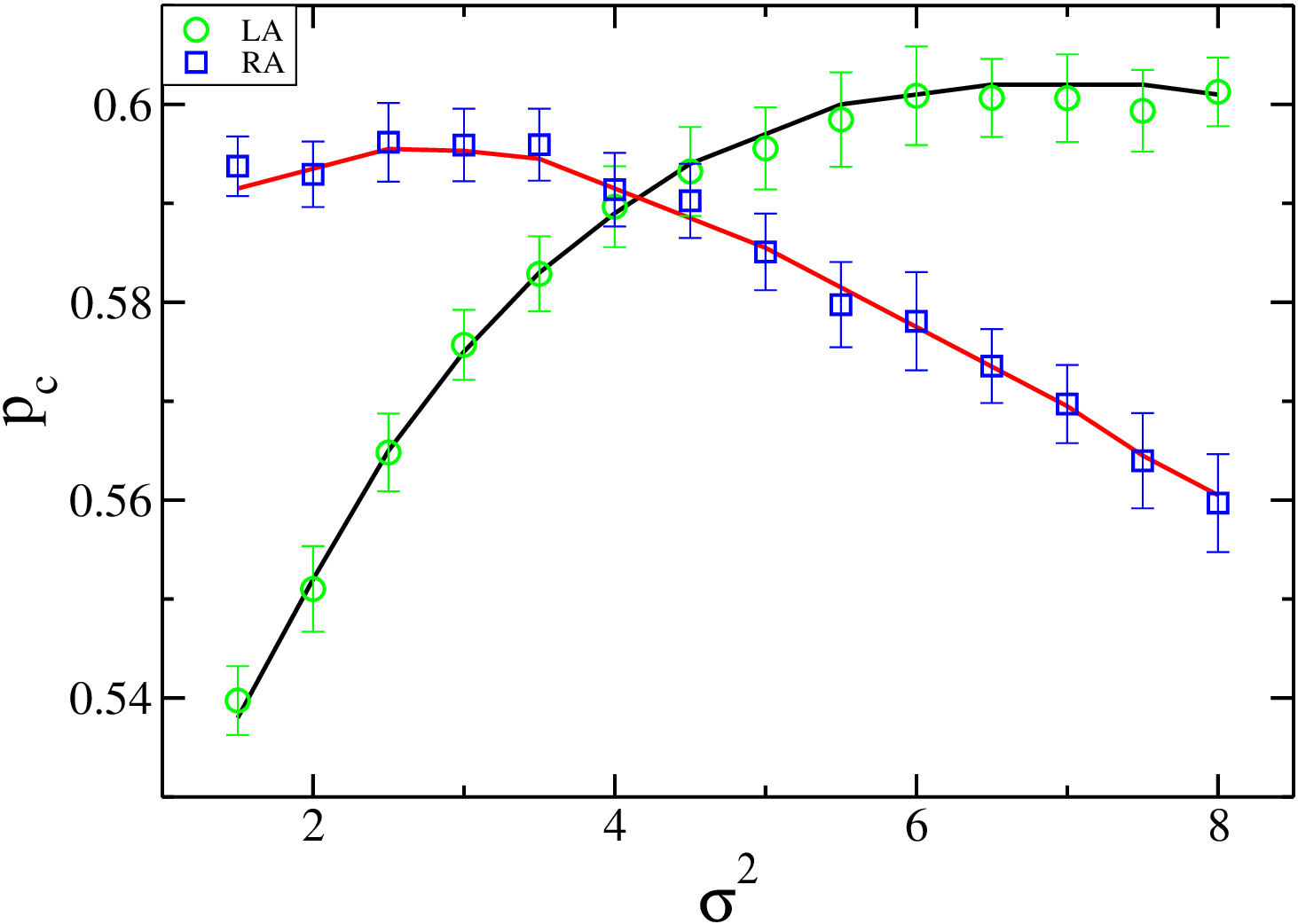}
\caption{\label{Fig10}(Color online) Percolation thresholds $p_c$ as a
  function of $\sigma^2$ of the fully interdependent Gaussian networks
  under LA and RA with $\mu=4$. Here solid lines are theoretical
  predictions, from Eq.~(\ref{tp_c}) for RA (red line) and similarly for
  LA (black line) and symbols ($\square$ for RA and $\bigcirc$ for LA)
  with error bars are simulation results with network size of $N=10^4$
  nodes, where averages and standard deviations are taken from 20
  realizations. It is seen here that as $\sigma^2$ increases $p_c(LA)$
  increases and $p_c(RA)$ has a tendency to decrease. As $\sigma^2$
  approaches the value of $\mu$, $p_c(LA)\approx p_c(RA)$, which is
  manifested by the intersection point shown here.}
\end{figure}

\subsubsection{Fully interdependent networks with Gaussian degree distribution}

We construct two fully interdependent networks in which the degrees in
each network follow the same Gaussian distribution and carry out a RA on
one of the networks to initiate a cascading failure process that will
continue until it reaches a steady state. We repeat the action, but this
time using a LA.  Figure~\ref{Fig9} shows the sizes of the giant
component $P_{\infty}(p)$ as a function of the occupation probability
$p$ under both LA and RA. Note that simulation results and the
theoretical results obtained from Eq.~(\ref{MCGG}) agree. 
 When $\mu=4$ and $\sigma^2=2$
the system is more fragile under LA than under RA with $p_c(LA)<
p_c(RA)$, and the giant components exhibit different behaviors.

If we fix $\mu$, when the Gaussian distribution gets broader, i.e., when
$\sigma$ increases, analogous to what we find in a single Gaussian
network, the critical $p_c$ behavior of the system differs under LA from
that under RA. Figure~\ref{Fig10} shows the effect of $\sigma$ on $p_c$
in the fully interdependent Gaussian networks. When $\sigma^2<\mu$,
$p_c(LA)<p_c(RA)$, and the opposite occurs when $\sigma^2>\mu$. The
intersection point in Fig.~\ref{Fig10} is located near
$\sigma^2\approx\mu$, similar to that in Poisson distribution
networks. Thus the system behaves the same under LA as under RA,
confirming the results presented in the previous subsection.
Note that our results show that in both attack scenarios, the interdependency
 of the system makes it much more vulnerable to RA and LA compared 
 to single networks (compare Fig.~\ref{Fig10} to Fig.~\ref{Fig5}).

\section{Conclusions}

In summary, we show that a LA on interdependent networks can be mapped
to a RA problem by transforming the network under initial attack. We
also show how the broadness of the degree distribution affects the
robustness of networks against RA and LA respectively. We show that, in
general, as the degree distribution broadens the network becomes more
vulnerable to LA than RA. This finding holds for both single networks
and interdependent networks.

\section*{Acknowledgments}

We wish to thank ONR, DTRA, NSF, the European MULTIPLEX, CONGAS and LINC
projects, DFG, the Next Generation Infrastructure (Bsik) and the Israel
Science Foundation for financial support. We also thank the FOC program
of the European Union for support.

\bibliographystyle{apsrev}

\bibliography{References}

\end{document}